\documentclass[twocolumn,showpacs,prl]{revtex4}
\usepackage{dcolumn}
\usepackage{graphicx}
\usepackage{graphicx,amssymb,amsmath,revsymb}
\usepackage{natbib}
\begin{document}


\title{Deformation of crosslinked semiflexible polymer networks}

\author{David A. Head$^{1,2}$}

\author{Alex J. Levine$^{2,3}$}

\author{F.C. MacKintosh$^{1,2}$}

\affiliation{$^{1}$Division of Physics \& Astronomy, Vrije Universiteit 
1081
  HV Amsterdam, The Netherlands}
\affiliation{$^{2}$The Kavli Institute for Theoretical
  Physics, University of California, Santa Barbara CA 93106, USA}
\affiliation{$^{3}$Department of Physics, University of Massachusetts,
Amherst MA 01060, USA}

\date{\today}

\begin{abstract}
Networks of filamentous proteins play a crucial role in cell mechanics. 
These
\emph{cytoskeletal} networks, together with various crosslinking and other 
associated proteins
largely determine the (visco)elastic response of cells. In this letter we 
study 
a model system of
crosslinked, stiff filaments in order to explore the connection
between the microstructure under strain and the macroscopic response
of cytoskeletal networks.
We find two distinct regimes as a function 
primarily of crosslink density and filament rigidity: one characterized by affine 
deformation and one by non-affine deformation. 
We characterize the crossover between these two. 
\end{abstract}


\maketitle

The study of biopolymer networks and gels lies at the heart of the
understanding of the mechanical properties of the cytoplasm since the
mechanical rigidity of the intracellular material is largely governed by
the \emph{cytoskeleton}, a complex network of filamentous proteins, 
cross links, and other associated proteins \cite{AlbertsEtc}. 
A key player in this cytoskeleton is F-actin, 
which exhibits significant rigidity on the cellular scale. 
The material properties of such
{\em semi-flexible} polymer networks also pose
complex and current problems in polymer physics.
Many of the most basic questions concerning these common and 
important networks, 
such as how they deform under stress, remain unanswered. 
The bending rigidity of such polymers introduces a
new microscopic elastic parameter that can have consequences for
the macroscopic elastic coefficients of the bulk, semi-flexible
gel. This changes the traditional rubber
elasticity model for the bulk properties of gels composed of
cross-linked flexible polymers.

In this letter, we examine a simple model for 
crosslinked rods that not only allows us to quantitatively test
the relationship between the microscopic and macroscopic elastic
coefficients of a randomly cross-linked network, but also sheds light
on the intimately related issue of the spatial distribution
of the network strain. 
Among the most fundamental properties of polymer networks
is the way in which they deform under stress. 
Since the classical theories of rubber elasticity \cite{Rubinstein}
it has been suggested that this deformation is \emph{affine}, 
\emph{i.e.}, that the strain is uniform, as for a sheared 
Newtonian fluid at low Reynolds number.
While this leads to relatively good agreement 
with experiments for rubber-like networks, much evidence and continuing
efforts concern systematic deviations that point 
to non-affine network strains at a microscopic scale.

The assumption that the deformation field is affine down to length
scales comparable to the smallest microscopic scales in the
material is a great simplification that allows one to construct 
quantitative theories relating the macroscopic elastic constants of a
gel to the microscopic properties of its constituent polymers. The
validity of the affine approximation for this class of semiflexible 
polymer materials has, however,
been the subject of some debate \cite{fcm1,frey,satcher}.
Whether the deformation field is
affine or not depends, of course, on length scale; clearly at the scale of 
the entire
sample, all deformations are trivially affine when subject to simple 
shear. We explore whether
this self-averaging property of the deformation field extends to
shorter length scales for semiflexible networks.
We show that the degree of non-affine strain is a function of length
scale and degree of crosslinking.
Specifically, we find that the range of non-affine strain can extend well
beyond the \emph{mesh size}, or \emph{correlation length}, 
the typical separation between filaments \cite{DeGennes}.
This occurs near the point of rigidity percolation for the network,
or for highly flexible filaments.
This results in elastic moduli governed 
primarily by bending of filaments under non-affine strain, consistent with 
\cite{frey,satcher}.
In contrast, we find that these networks become
increasingly affine, even down to the smallest scales of the network, \emph{e.g.} the mesh size, at high
cross link density, high molecular weight, or for rigid filaments.
Here, we find that the bulk elastic moduli converge to those
predicted from affine theory \cite{fcm1}. We also quantify the degree of non-affine strain and
show that this is, indeed, dependent on the length scale. 

As we focus on the
zero-frequency, or static properties of the system we may ignore
the complexities of the network-solvent interaction.  To isolate
the importance of semi-flexibility upon network properties, we
also ignore the complex, nonlinear response of the individual
F-actin filaments \cite{MarkoSiggia,fcm1,frey}. 
We instead study in detail the dependence of
the bulk shear modulus and Young's modulus of the material upon
the cross-link density of the polymer gel as well as the bending
and extension moduli of the individual filaments.
We model the network filaments via the Hamilitonian per unit
length ($\delta s$) for a filament
\begin{equation}
\frac{\delta{\cal H}}{\delta s} = \frac{\mu}{2} \left(
\frac{\delta l}{\delta s} \right)^{2} + \frac{\kappa}{2} \left(
\frac{\delta\theta}{\delta s} \right)^{2}. \label{e:hamiltonian}
\end{equation}
The first term takes into account the extensional
deformation of the filament $\delta l (s)$ as a function of
arc-length, $s$ with
modulus $\mu$.
The second term 
determines the energy stored in the filament due to bending: the
local tangent of the filament makes an angle $\theta (s)$ with
respect to the $\hat{x}$-axis and the  bending modulus of the
filament is $\kappa$. Note that both of these terms are quadratic. 
We do not explore nonlinearities,
such as buckling when compressed beyond the Euler instability,
at the scale of individual filaments.
While such nonlinear effects 
are expected at increasing strains and finite temperature \cite{fcm1}, 
we seek to understand here the fundamental
properties of semi-flexible networks and thus address only the
linearized version of the problem. 
We note, however, that thermal fluctuations can result in an effective modulus $
\mu\sim \kappa^2/(kT l^3)$ for a segment of length $l$. Thus, we consider $\mu$ 
and $\kappa$ to be independent parameters, even though they are both determined 
by single filament elastic properties and geometry at $T=0$ \cite{LL}. 
Clearly the full exploration of these
networks at finite temperature presents an interesting challenge;
understanding the $T=0$ mechanical properties
of these networks, however, is the requisite first step towards
this more ambitious program. In addition, we will point out
specific aspects of our results that are likely to depend
critically on our zero-temperature assumption. After all, it is
now understood that for some mechanical properties of networks,
the zero temperature presents a singular limit as in the case of
rigidity percolation.

Real F-actin networks formed from an actin monomer solution have a
complex geometry, arising from the dynamic growth and branching of
filaments~\cite{AlbertsEtc,real_networks}. For computational efficiency, we
ignore such complications and consider static, isotropic networks
of monodisperse filaments of length $L$. Each filament is
represented by a line segment deposited with random position and
orientation to a two--dimensional rectangular shear cell.
Intersections are identified as permanent, freely--rotating cross
links, to mimic {\em e.g.} the attachments of double--headed
myosin molecules in an ATP--deficient
solution~\cite{active_exps}.
The mean
distance between cross links is $l_{\rm c}$\, as measured along a
filament. Deposition continues until the desired cross link density
$L/l_{\rm c}$ has been reached.

The network is represented by the set of mobile nodes
$\{{\bf x}_{i}\}$ consisting of all cross links and midpoints
between cross links (the latter so as to
include the dominant bending mode).
The total system energy ${\cal H}(\{ {\bf x}_{i} \})$ is
then expressed in terms of the $\{{\bf x}_{i}\}$
using a discrete version of~(\ref{e:hamiltonian}).
Within our linearized scheme, this ${\cal H}(\{ {\bf x}_{i} \})$ is
a high--dimensional paraboloid with a unique global
minimum, corresponding to the state of mechanical
equilibrium at $T=0$.
For the initially unstressed network, this minimum corresponds
to zero deformation.
Depending on whether we wish to measure the shear modulus $G$ or
the Young's modulus $Y$, a shear or uniaxial strain $\gamma$ is
applied across the periodic boundaries in a Lees--Edwards
manner.
This moves the global minimum to a new, non--trivial position,
which we numerically find using the
preconditioned conjugate gradient method.
The stored energy per unit area can then be calculated,
which is $\gamma^{2}/2$ times $G$ or $Y$~\cite{LL}
within our linear approximation,
and hence the network modulus can be extracted.
This procedure is repeated
for different network realizations until a reliable estimate of
$G$ or $Y$ has been attained.
Further simulation details can be found in~\cite{long}.

Apart from the system size,
there are three length scales in the problem:
two geometric lengths $L$ and $l_{\rm c}$,
and a third material length scale deriving from
the stretching and bending modulii,
$l_{\rm b}=\sqrt{\kappa/\mu}$.
On dimensional grounds, the modulii can be
written in the form
$G=\frac{\mu}{L}f\left(L,l_{\rm c},l_{\rm b}\right)$
(with a similar expression for $Y$),
provided that the system size is sufficiently
large that finite size effects can be ignored, as applies
to all of the results presented here.

\begin{figure}[htpb]
\centering
\includegraphics[width=8.5cm]{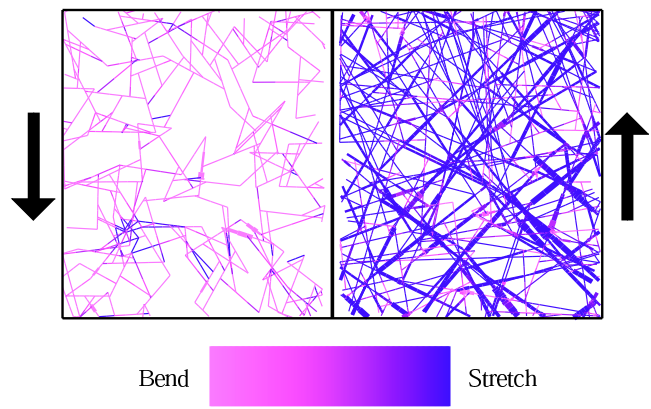}
\caption{{\em (Color online)}
Superimposed examples of the energy distribution
throughout networks of cross link densities
{\em (left)}~$L/l_{\rm c}\approx8.99$ and
{\em (right)}~$L/l_{\rm c}\approx46.77$
with $l_{\rm b}/L=0.006$
under the given shear strain.
The thickness of each segment
is proportional to the energy density per unit length,
with a minimum so that all filaments are visible,
and the calibration bar shows what proportion of the total
energy is due to stretching.
}
\label{f:pictures}
\end{figure}

\begin{figure}[htpb]
\centering
\includegraphics[width=9cm]{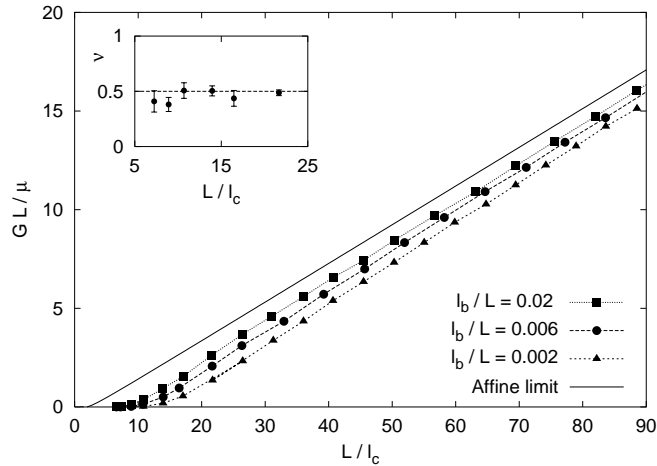}
\caption{Dimensionless shear modulus $GL/\mu$ versus the dimensionless cross
link density $L/l_{\rm c}$ for different $l_{\rm b}$,
demonstrating convergence to the affine solution at
high densities or molecular weights.
The error bars are no larger than the symbols. 
{\em (Inset)} The Poisson ratio
$\nu$ for $l_{\rm b}/L=0.006$. }
\label{f:modulii}
\end{figure}

Results are given in Fig.~\ref{f:modulii} for the shear modulus
$G$ and the Poisson ration $\nu=2Y/G-1$.
$G$ monotonically increases with the cross link
density $L/l_{\rm c}$ and the ratio $l_{\rm b}/L$.
Both $G$ and $Y$ simultaneously vanish at the rigidity
percolation transition
$[L/l_{c}]_{\rm crit}\approx5.9$, irrespective of $l_{b}$.
This is consistent with the more precise value
$[L/l_{\rm c}]_{\rm crit}\approx5.932$ found by
Latva--Kokko {\em et al.}~\cite{latva_kokko} using the
combinatorial ``pebble game'' method.
We find that $G$ and $Y$ scale near the
transition as $\sim(L/l_{\rm c}-[L/l_{\rm c}]_{\rm crit})^{f}$
with $f=3.0\pm0.2$, placing it in a distinct universality class
from both central--force rigidity and bond bending {\em without}
free rotation at cross links~\cite{rigidity}. The variation
of $\nu$ is more subtle if at all, varying from $\approx0.5$ for
high densities to $\approx0.35\pm0.1$ near the critical point
(for comparison, standard stability considerations require
$-1\leq\nu\leq1$ in 2D). 
Further details of the scaling behavior near the
transition will be presented elsewhere~\cite{long}.
We remark
that at finite temperatures $G$ will remain non--zero
above the {\em conductivity} percolation transition at
$L/l_{\rm c}\approx5.42$, but it is not clear if
this small--$G$ behavior is experimentally observable for the
macromolecules under consideration here.

Away from the critical point, $G$ decreases as the bending modulus
$\kappa$ decreases. Indeed, in the limit $\kappa\rightarrow0$, or
equivalently $l_{\rm b}\rightarrow0$, all filaments freely bend
and the model reduces to a random network of Hookean springs,
which is already known to have a vanishing $G$ for all finite
cross link densities~\cite{zero_G}. In the opposite limit
$l_{\rm b}\rightarrow\infty$, bending along a filament's length
becomes prohibitively expensive and response becomes dominated by
stretching modes. The same is true of the limit $L/l_{\rm
c}\rightarrow\infty$ with $l_{\rm b}/L$ fixed, since bending modes
of wavelength greater than $l_{\rm c}$ require an {\em area} of
network to twist rather than a single filament, and are
suppressed. This transition from low $L/l_{\rm c}$,
bending--dominated networks to high $L/l_{\rm c}$,
stretching--dominated ones  can be clearly seen 
Fig.~\ref{f:pictures}.

Pure, affine shear, being a combination of rotation and extension, induces 
only 
stretching and compression of filaments. Furthermore, networks dominated 
by 
stretching modes must be approximately affine.
This follows from the observation that, far above the rigidity
transition, there can be no orientational disorder without a
corresponding cost in bending energy. It is straightforward to
derive analytic expressions for the modulii under an affine
strain. A rod of length $L$ lying at an angle $\theta$ to the
$x$--axis will undergo a relative change in length $\delta
L/L=\gamma_{\rm xy}\sin\theta\cos\theta$ in response to an affine
strain field $\gamma_{\rm xy}$\,. According to
(\ref{e:hamiltonian}) with $\delta\theta=0$, the energy cost
(after uniformly averaging over all angles $\theta$) is
$\delta{\cal H}=\mu L\gamma_{\rm xy}^{2}/16$. To calculate $G$, we
need to express the number of rods per unit area $N$ as a function
of $L$ and $l_{\rm c}$. The exact expression is easy to derive,
but for current purposes it is sufficient to use the approximate
relation $L/l_{\rm c}\approx(\alpha-1)/(1-2/\alpha)$ with
$\alpha=2L^{2}N/\pi$, valid for $L/l_{\rm c}\approx5$ and greater
(note that the full expression is monotonic). A further correction
removes the dangling ends of the rods by renormalizing the rod
lengths to $L-2l_{\rm c}$. Then using $G=2{\cal H}/\gamma_{\rm
xy}^{2}$,
\begin{equation}
G_{\rm affine}=\frac{\pi}{16}\frac{\mu}{L}
\left(
\frac{L}{l_{\rm c}}
+2\frac{l_{\rm c}}{L}
-3
\right)
\label{e:G_affine}
\end{equation}
so that $G\sim(\pi/16)\mu/l_{\rm c}$ as $L/l_{\rm
c}\rightarrow\infty$. Expression (\ref{e:G_affine}) is plotted in
Fig.~\ref{f:modulii} and gives a reasonable approximation to the
data, with the agreement improving as $L/l_{\rm c}$ increases. The
same calculations can be repeated to give $Y_{\rm affine}=3G_{\rm
affine}$ and hence a Poisson ratio $\nu_{\rm affine}=0.5$, which
is also plotted in the figure. Perhaps surprisingly, we find 
remarkably good agreement between the measured
$\nu$ and the affine value, even close
to the rigidity transition, where the affine approximation fails.

\begin{figure}[htpb]
\centering
\includegraphics[width=9cm]{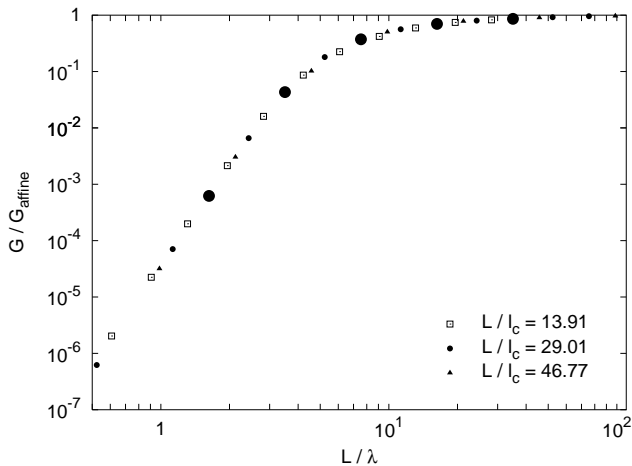}
\caption{The master curve of $G/G_{\rm affine}$
plotted against $L/\lambda$ with
$\lambda=\sqrt[3]{l_{\rm c}^{4}/l_{\rm b}}$.
The enlarged points for $L/l_{\rm c}\approx29.09$
correspond to the same parameter values as in Fig.~\ref{f:affinity}.}
\label{f:master}
\end{figure}

We observe that there is an apparent crossover from a bending-dominated, 
non-affine 
regime for either high molecular weight ($L$) or for high density. The 
natural measure
for this is the ratio of filament length to average distance between 
crosslinks. 
This can be understood by the fact that, unlike networks of flexible 
polymers, where 
segments along a single polymer between crosslinks appear to behave as 
effectively independent
network strands, the segments of semiflexible filaments between crosslinks 
act in series.
Thus, segments far from free ends are forced to deform nearly affinely by 
the many 
constraints on their neighboring segments imposed by crosslinks. This 
suggests a 
physical picture in which non-affine deformations are primarily associated 
with 
less constrained free ends. Let $\lambda$ denote the range of such 
non-affine 
regions near the filament ends. We expect this length 
$\lambda\left(l_c,l_b\right)$ 
to be a function of the local 
density of filaments (measured by $l_c$, the distance between crosslinks) 
and the
material length $l_b$. Then, what determines the degree to which the 
network is affine or not 
is the relative size of the affine to non-affine regions, or the ratio 
$L/\lambda$. 
We find that, sufficiently far above the transition,
$G/G_{\rm affine}$ collapses to a single master curve
with the empirical choice $\lambda=\sqrt[3]{l_c^4/l_b}$, 
as shown in Fig.~\ref{f:master}
(the origin of this length scale is
discussed elsewhere~\cite{long}).
This confirms the existence of a fundamental length scale $\lambda$ for 
non-affine deformations
along the filament backbone, as well as two physically distinct regimes: 
non-affine behavior for 
$L\alt\lambda$ and affine behavior for $L\agt\lambda$.  
In the first of these, we note that the modulus depends only on $\kappa$ 
and not on $\mu$, 
signalling a bending dominated regime, as predicted by Kroy and 
Frey~\cite{frey,frey2}.
Similar results have been found
independently~\cite{frey3}.
Note that the crossover can be reached at any fixed density
by varying $l_{\rm b}$ alone, and thus represents
distinct physics from the percolation transition,
which exists only at a specific density.
Hence we use absolute density,
rather than its value relative to the transition,
in our scaling function.

\begin{figure}[htpb]
\centering
\includegraphics[width=9cm]{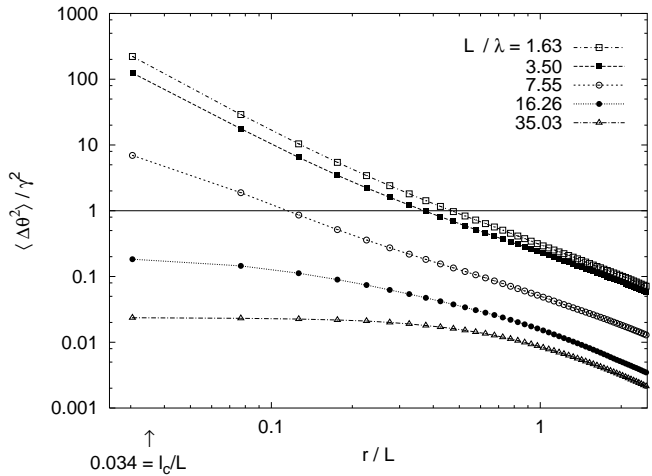}
\caption{Plot of the affinity measure
$\langle\Delta\theta^{2}(r)\rangle$
normalised to the magnitude of the imposed strain $\gamma$
against distance~$r/L$, for different $l_{\rm b}/L$.
The value of $r$ corresponding to the mean distance
between crosslinks $l_{\rm c}$ is also indicated,
as is the solid line
$\frac{1}{\gamma^{2}}\langle\Delta\theta^{2}(r)\rangle=1$,
which separates affine from non--affine networks
to with an order of magnitude.
In all cases, $L/l_{\rm c}\approx29.1$
and the system size was $W=\frac{15}{2}L$.
}
\label{f:affinity}
\end{figure}

Our findings demonstrate that the degree of affinity of the
deformation field does depend on the length scale on which one looks. This 
is shown 
in Fig.~\ref{f:affinity},
where the quantity 
$\langle\Delta\theta^{2}(r)\rangle$
is plotted against $r$ for different values of $L/\lambda$.
$\langle\Delta\theta^{2}(r)\rangle$ is the square
deviation of the angle of rotation $\theta$ between
two points separated by a distance $r$,
relative to the affine equivalent.
This monotonically decreases for increasing $r$, suggesting
that the deformation appears more affine
when viewed on larger length scales.
Furthermore, the deviation from affinity at the mesh scale
$r=l_{\rm c}$ is small for networks with
$G\approx G_{\rm affine}$,
and large for those with $G\ll G_{\rm affine}$, as seen by
comparing Figs.~\ref{f:affinity} and \ref{f:master}.
Thus we can reiterate the main findings of our work:
(1)~there are two qualitatively distinct regimes, one affine and the other non-affine;
(2)~the physics of the crossover between these is distinct from the rigidity transition;
and~(3) the crossover is governed by a new length scale
$\lambda$, where, \emph{e.g.}, affine behavior 
is seen for filament lengths a few times this length.
Since this length is expected to be of order 
the distance between crosslinks, real networks can be in either
regime depending on the length distribution.

AJL would like to acknowledge the hospitality of the Vrije
Universiteit. DAH was partly
funded by a European Community Marie Curie Fellowship. This work is 
supported in part by the National Science
Foundation under Grant Nos. DMR98-70785 and PHY99-07949.


\end{document}